\documentstyle[12pt,amsmath]{article}
\begin{document}
\begin{titlepage}
\quad\\
\vspace{1.8cm}
\begin{center}
{\bf\LARGE Crossover quintessence and cosmological history of}\\
\bigskip
{\bf\LARGE fundamental ``constants''}\\
\vspace{1cm}
C. Wetterich\footnote{e-mail: C.Wetterich@thphys.uni-heidelberg.de}\\
\bigskip
Institut  f\"ur Theoretische Physik\\
Universit\"at Heidelberg\\
Philosophenweg 16, D-69120 Heidelberg\\
\vspace{3cm}
\begin{abstract}
Crossover quintessence predicts that the time variation of
fundamental couplings is substantially faster at
redshift $z\approx 2$ than over the past few billion years. This could reconcile the
reported time variation of the fine structure constant from quasar absorption lines with
severe bounds from archeo-nuclear physics and high precision tests of the
equivalence principle. We present a model that is consistent with all present data on time
varying couplings, tests of the equivalence principle and cosmology.
\end{abstract}
\end{center}\end{titlepage}
\newpage

Quintessence or a time varying dark energy can be associated with a scalar field whose
cosmological expectation value varies during the recent history of the universe \cite{CW1}.
Generically, this scalar field - the cosmon - may also couple to  matter and radiation. As
a consequence, the values of ``fundamental constants'' like the fine structure constant or
the ratio between the nucleon mass and the Planck mass depend on the value of the cosmon field
and therefore on cosmological time \cite{CW1}-\cite{Dvali}. Recently, a time variation
of the fine structure constant has been reported \cite{Webb} from the observation of quasar
absorption lines (QSO) at redshift $z\approx 2$. A typical value corresponds to
\begin{equation}\label{1}
\frac{\Delta\alpha_{em}(z=2)}{\alpha_{em}}=-7\cdot10^{-6}.
\end{equation}
Similar observations infer \footnote{A recent collection of results on
$\Delta\alpha_{em}$ as well as the time variation of other fundamental constants can be
found in \cite{Uzan}.} a substantially smaller value of $|\Delta\alpha_{em}|$ at lower
redshift $z\le 0.7$. Furthermore, from the Oklo natural nuclear reactor one
obtains a typical bound \cite{Oklo},
\cite{Re}
\begin{equation}\label{2}
\frac{|\Delta\alpha_{em}(z=0.13)|}{\alpha_{em}}<10^{-7}
\end{equation}
and analysis of the decay rate $Re^{187}\rightarrow Os^{187}$ from meteorites
restricts \cite{Re}
\begin{equation}\label{2a}
\frac{|\Delta\alpha_{em}(z=0.45)|}{\alpha_{em}}<3\cdot10^{-7}.
\end{equation}

As an obvious question one may ask if an increase of $|\Delta\alpha_{em}|$ by almost two
orders of magnitude between $z=0.13$ and $z=2$
can reasonably be explained by quintessence models. In this note we
demonstrate that this can indeed be the case in a class of models of ``crossover
quintessence'' proposed recently \cite{CWCF}. This contrasts to constant rates
$\partial\Delta\alpha_{em}/\partial z=const$. or $\partial\Delta\alpha_{em}/\partial t=const.$
for which the values and bounds (\ref{1})-(\ref{2a}) clearly are in discrepancy. The
deviation from constant rates also affects strongly the comparison of eq. (\ref{1}) with
bounds from nuclear synthesis \cite{BBN},\cite{Av} and the cosmic microwave
background (CMB) \cite{Av}.

We are aware that the bound from the Oklo natural reactor is subject to substantial
uncertainties due to possible cancellations between effects from the variation of
$\alpha_{em}$ and other ``fundamental constants'' like the mass ratio between pion
and nucleon mass $m_{\pi}/m_n$ or the weak interaction rates. Also the
QSO observations need further verification and investigation of systematics.
Nevertheless, it is well conceivable that we can
get a reasonable ``measurement'' of the function $\Delta\alpha_{em}(z)$ in a not too
distant future. Furthermore, the derivative $\partial\Delta\alpha_{em}(z)/\partial z$ at
$z=0$ can be related \footnote{See \cite{Dvali} for a different relation
which gives a much smaller violation of the equivalence principle.}
\cite{CWTV} to precision tests of the equivalence principle \cite{EP}.
The aim of this letter is to present a sample computation how the knowledge of
$\Delta\alpha_{em}(z)$ can be used to constrain models of quintessence. For this purpose we
will use the value (\ref{1}) as a ``benchmark value'' which fixes the strength of the cosmon
coupling to matter and radiation. The time history $\Delta\alpha_{em}(z)$ can then be
related to the time history of quintessence.

In particular, we may investigate the ratio
\begin{equation}\label{AA1}
R=\Delta\alpha_{em}(z=0.13)/\Delta\alpha_{em}(z=2).
\end{equation}
A bound $R<1/50$ strongly favors \cite{CWTV}
quintessence with a time varying equation of state $w_h=p_h/\rho_h$, where the value of
$(1+w_h)$ at present is substantially smaller than for $z=2$. A logarithmic dependence of
$\Delta\alpha_{em}$ on the cosmon field $\chi$ leads to a bound
for the present equation of state \cite{CWTV}
\begin{equation}\label{AA2}
w^{(0)}_h<-0.9.
\end{equation}
In this note we investigate
how the detailed dependence of $\alpha_{em}$ on $\chi$ influences the time history of
$\Delta\alpha_{em}(z)$. We argue that reasonable $\beta$-functions for the ``running'' of
a grand unified coupling with $\ln\chi$ may lead to nonlinearities in $\alpha_{em}(\ln\chi)$.
Those are quantitatively important, influencing $R$
within a factor two. The overall picture remains rather solid,
however: a small value of $|R|$ requires a quintessence model where the evolution of the
cosmon field has considerably slowed down in the recent history as compared to a redshift
$z\approx 2$. This feature is characteristic for crossover quintessence and will not be
shared by arbitrary quintessence models. In particular, a constant equation of state with
$w_h$ independent of $z$ over a range $0<z<3$ will have severe difficulties to explain
$R<0.1$. This demonstrates how a measurement of $\Delta\alpha_{em}(z)$ could become an
important ingredient for the determination of the nature of dark energy.

Crossover quintessence (CQ) can be characterized by a recent increase of the fraction in
homogeneous dark energy $\Omega_h=\rho_h/\rho_{cr}$ from a small but not negligible value
for early cosmology to a value $\Omega_h\approx0.7$ today. A small early value (say
$\Omega_h=0.01)$ typically results from a cosmic attractor solution \cite{CW1,Rat,Att}
(``tracker solution'') independently of the detailed initial conditions. For such models
of ``early quintessence'' the dark energy is diluted in early cosmology at the same pace as
the dominant energy component. Dark energy and radiation or matter have therefore always been of a comparable
magnitude and excessive ``fine tuning'' can be avoided \cite{CW1,Heb}. The amount of early
quintessence is accessible to observation by nucleosynthesis \cite{CW1,NS},
CMB \cite{Doran} or structure formation \cite{FJ,Schwindt}.

The crossover to domination by dark energy must have happened after the formation of
structure. On the level of the equation of state CQ
corresponds to a crossover from a value $w_h \approx 1/3$ during the radiation dominated
universe to a substantially negative value for $z<0.5$. During the crossover period the time
variation of the cosmon field remains substantial - this  may extend to $z\approx 1.5$
and be the origin of the QSO observation (\ref{1}). More recently, towards the end of the
crossover period, the equation of state may
have approached the value $w_h\approx-1$, corresponding to a suppressed cosmon kinetic energy $T=(1+w_h)
\rho_h/2$ and therefore a slowing down of the evolution of the cosmon. For a present value
$1+w^{(0)}_h\ll 1$ the evolution of the cosmon and the fundamental couplings has almost
stopped, explaining the small bound (\ref{2}). We note that the fraction of energy density
in matter, $\Omega_m$, decreases rapidly once $w_h$ has come close to $-1$ at the end
of the crossover. The observed present value $\Omega^{(0)}_m\approx0.3$
therefore indicates that
the effective ``stop'' in the cosmon evolution should have happened rather recently - a value $z\approx 1$ for the drop in the
variation rate of the fundamental couplings seems rather natural in this respect.

In this letter we want to give a specific example for this scenario. For this purpose
we elaborate on a recent proposal \cite{CWCF} that CQ is connected to the
existence of a conformal fixed point in a fundamental theory. The vicinity to a fixed
point can give a natural explanation why the time variation of fundamental couplings is
only a tiny effect. Our model resembles in this respect the ``runaway dilaton'' in string
theories \cite{Dam}. We want to demonstrate here the existence of a model which is
compatible with (\ref{1})-(\ref{2a}) rather than to advocate the detailed choice of the
$\beta$-functions for the running couplings below.

Let us assume that for low momenta
$(q^2\ll\chi^2)$ the yet unknown unified theory results in an effective
grand unified model in four dimensions.
There the coupling of the cosmon field $\chi$ to gravity and the gauge fields is described
by an effective action
\begin{eqnarray}\label{3}
S&=&\int d^4x\sqrt{g}\left\{-\frac{1}{2}\chi^2R+\frac{1}{2}(\delta(\chi)-6)\partial^{\mu}
\chi\partial_{\mu}\chi\right.\nonumber\\
&&\left.+m^2\chi^2+\frac{Z_F(\chi)}{4}F^{\mu\nu}F_{\mu\nu}\right\}.
\end{eqnarray}
We note that the value of the cosmon field $\chi$ plays the role of a dynamical mass
scale in the fundamental theory. Typically, it could be associated with the
characteristic scale for the transition of a higher dimensional theory to an effective
four dimensional ``low energy'' theory. We have normalized $\chi$ such that its
present value equals the reduced Planck mass $\bar{M},\bar{M}^2=1/(8\pi G_N)$.
Typical particle masses are $\sim \chi$. Up to the $\chi$-dependence of the running
dimensionless couplings $\delta$ and $Z_F$ and the explicit mass term $m^2\chi^2$
the effective action is dilatation invariant. The global dilatation symmetry
$\chi\rightarrow\alpha\chi$ is spontaneously broken by any nonzero cosmological
value of $\chi$.

The running of the effective grand unified gauge coupling
$g^2_X=\bar{g}^{2}/Z_F$ is determined
by the $\chi$-dependence of $Z_F$. The assumed fixed point behavior for
$\alpha_X=g^2_X/4\pi$ is cast into the form
\begin{equation}\label{4}
\frac{\partial\alpha_X}{\partial\ln\chi}=\beta_\alpha=b_2\alpha_X-b_4\alpha^2_X-b_6\alpha^3_X
\left(6-\frac{\delta}{1+J_{\alpha}\delta}\right).
\end{equation}
We take here a small value of $|b_6|$ and impose a fixed point value
\begin{equation}\label{5}
\alpha_{X,*}\approx\frac{b_2}{b_4}\approx\frac{1}{40}
\end{equation}
as extrapolated from the unification of the observed gauge couplings of the standard model.
The term $\sim b_6$ accounts for the dependence of the fixed point on the value of the other
dimensionless coupling $\delta$.
As mentioned above, we do not assign any particular significance to the specific form
of $\beta_\alpha$. We only assume the existence of a fixed point in the approximation
of constant $\delta$ which depends slightly on the value of $\delta$. In our case the
approximate shift in the location of the fixed point between $\delta=0$ and
$\delta=\infty$ amounts to \footnote{We have introduced $J_\alpha$ such that the shift
remains finite for $\delta\rightarrow\infty$.} $\Delta\alpha_{X,*}=b_6\alpha^3_{X,*}/
(b_2J_\alpha)$. Many other possible shapes of $\beta_\alpha$ will lead to similar
conclusions provided that one can give a reason why the variation of $\alpha_X$ is small.

For $b_2>0$ and $\alpha_X(\chi)>\alpha_{X,*}$
the grand unified gauge coupling increases with decreasing
$\chi$. Similar to QCD it grows large at some nonperturbative scale (``confinement scale'')
that we associate with $m$. Our model therefore has an ``infrared scale'' $m$ which is
generated by dimensional transmutation from the running of the grand unified
gauge coupling. We assume
that the non-perturbative scale
$m$ determines the mass of the cosmon. Neglecting the effects of the explicit dilatation
symmetry breaking reflected by $m$ the cosmon potential would be flat.

On the other hand the gauge coupling runs towards its ultraviolet fixed point
(\ref{5}) as $\chi$ increases.
At the fixed point the effective action would have an effective dilatation symmetry if the
running of $\delta(\chi)$ could be neglected. We will assume, however, that the running of
$\delta$ is unstable towards large $\chi$ and postulate $(E>0)$
\begin{equation}\label{6}
\frac{\partial\delta}{\partial\ln\chi}=\beta_\delta(\delta)=
\frac{E\delta^2}{1+J_{\delta}\delta}.
\end{equation}
For small enough $\delta_i=\delta(\chi=m)$ the dimensionless coupling $\delta$ stays small
for a large range $m<\chi\le\chi_c$. For $\chi\gg m$ we may neglect the small mass $m$ for the
evolution of the dimensionless couplings. The evolution in this range is then governed by the
vicinity to a conformal fixed point \cite{CWCF} at
$\delta=\delta_*=0~,~\alpha_X=\alpha_{X,*}$.
In the same approximation our model corresponds to a flat direction \footnote{In string
theories such flat directions are often called ``moduli''.  In presence of additional matter
fields, e. g. scalars and fermions, the flat direction in the space of scalar fields
corresponds to constant ratios of all masses \cite{CWCF}, up to the small effects from
varying couplings discussed in this note. The flat direction is consistent with the conformal
symmetry, but not required by it. See \cite{CWCF} for a more detailed motivation of a flat
direction in the presence of quantum effects.} in the effective cosmon potential.
All cosmon interactions are therefore given by its
derivative coupling to the graviton and gauge fields.

Nevertheless, as $\chi$ increases further an effective ``ultraviolet scale'' $\chi_c$
characterizes the region where $\delta$ grows large according to the approximate solution
\footnote{We take a small value $J_{\delta}=0.05$ such that only the behavior for large
$\delta$ is affected by $J_{\delta}$. The cosmology depends only very little on
$J_{\delta}$ in this case.}
\begin{equation}\label{7}
\delta(\chi)\approx\frac{1}{E\ln(\chi_c/\chi)}.
\end{equation}
The ratio between the ultraviolet and infrared scales turns out exponentially huge
\begin{equation}\label{8}
\frac{\chi_c}{m}\approx\exp\left(\frac{1}{E\delta_i}\right).
\end{equation}
At the crossover scale $\chi_c$ the flow of the couplings witnesses a crossover from
the vicinity of the conformal fixed point (small $\delta$) to another regime
for large $\delta$ - in our case a fixed point for the effective cosmon-gravity coupling
$\delta^{-1}$ at $\delta^{-1}_*=0$. We will see that the crossover behavior
in the present epoch of the cosmological evolution precisely corresponds to the crossover
from small to large $\delta$ for $\chi$ in the vicinity of $\chi_c$. Realistic cosmology
obtains for $E\delta_i\approx 1/138$.

The cosmological dynamics of CQ is described in \cite{Heb,CWTV}.
For given $E$ and $J_{\delta}=0.05$ we solve the flow equation (\ref{6}). The ``initial value''
$\delta_i=\delta(\chi=m)$ is tuned such that at present $\Omega^{(0)}_h$ equals the value
extracted from observation. The function
$\delta(\chi)$ in the effective action (\ref{3}) is now fixed and we can derive the field
equations for the evolution of gravity and the cosmon in a homogeneous and isotropic universe
\cite{CWTV}. For simplicity we neglect the small influence of radiation and matter on the cosmon
field equation which results from $\partial Z_F/\partial\ln\chi\neq 0$. For a solution of the field
equations we start for
$\chi(t=0)=m$ with rather arbitrary values of $\dot{\chi}(0)$ and $\Omega_h(0)$.
The late time behavior will not depend on this.

In table 1 we present
a few characteristic cosmological quantities, namely the present Hubble parameter
$(h)$, the present fraction of dark energy, $\Omega^{(0)}_h$,
the present equation of state,
$w^{(0)}_h$, the age of the universe, $t^{(0)}$, the fraction of dark energy at last
scattering, $\Omega^{(ls)}_h$, the location of the third peak in the CMB anisotropies in
angular momentum space, $l_3$, as well as the normalization of the spectrum of density
fluctuations, $\sigma_8$, divided by the value which would obtain for a cosmological
constant with the same $\Omega^{(0)}_h$. Crossover quintessence with cosmological
parameters in this range is well compatible \cite{CDMSW} with the recent precision
measurements of the CMB-anisotropies by WMAP \cite{WMAP}.

\vspace{0.5cm}
\noindent
\label{table1}
\begin{tabular}{|c|c|c|c|c|c|c|c|}\hline
$E$&
$h$&
$\Omega^{(0)}_h$&
$w^{(0)}_h$&
$t^{(0)}/10^{10}yr$&
$\Omega^{(ls)}_h$&
$l_3$&
$\sigma_8/\sigma^{(\wedge)}_8$\\ \hline
$5$&
$0.66$&
$0.70$&
$-0.93$&
$1.37$&
$0.019$&
$796$&
$0.70$\\
$12$&
$0.71$&
$0.73$&
$-0.99$&
$1.35$&
$0.0083$&
$794$&
$0.85$\\ \hline
\end{tabular}

\vspace{0.5cm}
\noindent
Table 1: Characteristic cosmological quantities for two CQ models.\\
\\
We recall that smaller $h$ and larger $\Omega^{(0)}_h$ shift $l_3$ to larger values.
We conclude that for large enough $E$ our model is consistent with present cosmological
observations.

In a grand unified theory the value of the electromagnetic fine structure constant depends
on the gauge coupling at the unification scale, $\alpha_X$, the ratio between the weak scale
and the unification scale, $M_W/M_X$, and particular particle mass ratios like
$m_n/M_W$ or $m_e/M_W$ for the nucleons and electrons. Since the nucleon mass is determined
by the running of the strong gauge coupling similar dependencies arise for the ratio
$m_n/M_X$.

For our model we assume that the dominant effect is due to the variation of $\alpha_X$
\cite{CWTV} and consider the approximation of time-independent mass ratios
$M_X/\bar{M},M_W/m_n$, $M_W/m_e$ etc. The dependence of the fine
structure constant on redshift can then be related directly to the $\beta$-function for the
grand unified gauge coupling (\ref{4}). For
$\Delta\alpha_{em}(z)=\alpha_{em}(z)-\alpha_{em}(0)$ one finds
\begin{equation}\label{19}
\frac{\Delta\alpha_{em}(z)}{\alpha_{em}}=\frac{22\alpha_{em}}{7\alpha^2_X}\Delta\alpha_X(z).
\end{equation}
The same approximation yields \footnote{This relation differs from \cite{GUT}
where $M_W/M_X$ is kept fixed instead of fixed $m_n/M_W$ in the present work. The
quantitative difference is not very important, however.}
\begin{equation}\label{N1}
\frac{\Delta(m_n/\bar{M})}{(m_n/\bar{M})}=\frac{\pi}{11\alpha_{em}}
\frac{\Delta\alpha_{em}}{\alpha_{em}}=39.1\frac{\Delta\alpha_{em}}{\alpha_{em}}.
\end{equation}
Grand unification links the time variation of $\alpha_{em}$ with the time variation of
$m_n/\bar{M}$ or the effective Newton constant, $\dot{G}/G=2d\ln(m_n/\bar{M})/
dt=78.2\dot{\alpha}_{em}/\alpha_{em}$.

In terms of the variable $x=\ln a=-\ln(1+z)$ the time dependence of
$\alpha_X(x)$ obeys (cf. \cite{CWTV})
\begin{equation}\label{20}
\frac{\partial\alpha_X}{\partial x}=\beta_{\alpha}\frac{\partial\ln\chi}{\partial x}
=\frac{\beta_{\alpha}(\alpha_X,\delta)}{\sqrt{\delta}}
\sqrt{3\Omega_h(1+w_h)}.
\end{equation}
One observes that the $z$-dependence of $\Delta\alpha_{em}$ is particularly weak in the
region of large $\delta$ and $w_h$ close to $-1$. This will be our explanation why
$R$ is considerably smaller than expected
from a simple extrapolation linear in $z$. Crossover quintessence is precisely characterized
by large $\delta$ and $w_h\approx-1$ in the region of small $z$, whereas $\sqrt{1+w_h}
/\sqrt{\delta}$ is considerably larger for intermediate $z\approx 1.5-3$. One also observes
that for constant $(\Omega_h(1+w_h)/\delta)^{1/2}$ and $\beta_{\alpha}/\alpha$ the
dependence of $\Delta\alpha_{em}(z)$ is logarithmic in $1+z$. This is crucial for large
$z$ as for nucleosynthesis, where $\Delta\alpha_{em}$ turns always out to be much smaller
than expected from a linear extrapolation in redshift or time.

In eq. (\ref{20}) the factor $1+w_h$ is generic for all scalar models since it reflects
the kinetic energy and therefore the rate of change of the scalar field. The substantial
change in this factor is the main effect for a possible explanation of a small value of
$R$ by CQ. The factor $(\Omega_h/\delta)^{1/2}$ is constant for small $\Omega_h$ in our model.
It may decrease at late time if $\delta$ grows large since $\Omega_h<1$ remains bounded.
Finally, the details depend on the form of $\beta_{\alpha}$. In our model the fine
structure constant would decrease with time for $b_2>0$ and $b_6=0$, at a rate that would
not be observable for the value $b_2=0.2$ considered here. However, the influence of the
change of $\delta$ on the location of the fixed point $\alpha_{X,*}$ leads to an increase
of $\alpha_{em}$. As a consequence, the overall normalization of the time variation strongly
depends on $b_6$. As a benchmark we use eq. (\ref{1}). Therefore,
for given $b_2$ and $J_{\alpha}$ we fix $b_6$ such that
$\Delta\alpha_{em}(z=2)=-7\cdot10^{-6}$. The shape of $\Delta\alpha_{em}(z)$ depends
now on $b_2$ and $J_{\alpha}$ as well
as on on $E$. In table 2 we show the values of $\Delta\alpha_{em}$ at $z=0.13,0.45,1100$ and
$10^{10}$ for $b_2=0.2,J_{\alpha}=6$.
An investigation of other choices of $b_2$ and $J_{\alpha}$ (in a range $b_2>0.05$)
typically shows (for given $E$) a spread in the value of $R$ by a factor $\approx 2$.

\vspace{0.5cm}
\noindent
\label{table2a}
\begin{tabular}{|c|c|c|c|c||c|c|}\hline
E&
$z=0.13$&
$z=0.45$&
$z=1100$&
$z=10^{10}$&
$R$&
$\eta$ \\ \hline
$12$&
$-6\cdot 10^{-8}$&
$-2.6\cdot 10^{-7}$&
$-6.5\cdot 10^{-5}$&
$-7.8\cdot 10^{-5}$&
$0.0085$&
$5.1\cdot 10^{-14}$ \\ \hline
$5$&
$-1.7\cdot 10^{-7}$&
$-8.2\cdot 10^{-7}$&
$-4.2\cdot 10^{-5}$&
$-5.4\cdot 10^{-5}$&
$0.024$&
$4.4\cdot 10^{-14}$\\ \hline
\end{tabular}

\vspace{0.5cm}
\noindent
Table 2: Time variation of the fine structure constant $\Delta\alpha_{em}(z)/\alpha_{em}$ for
various redshifts $z$. The last two columns show $R$ (eq. (\ref{AA1})) and the differential
acceleration $\eta$.

Consistency with the Oklo bound (\ref{2}) and the Re-decay bound (\ref{2a}) is found
for large $E$. By looking at smaller values of $E$ it also becomes clear that a
precise knowledge of $\Delta\alpha_{em}(z)$ could be used as a probe for
the dynamics of quintessence. For example, it seems very difficult to achieve
$R\leq 1/50$ for small values of
$E$, quite independently of the precise form of $\beta_{\alpha}$. The present investigation
confirms that a small value of $R$ requires $(1+w^{(0)}_h)\ll 1$. This property of the
equation of state seems to be favored by the cosmological tests (cf. table 1). We conclude
that the bound (\ref{AA2})
can be justified within a very broad class of models if one requires $|R|<0.02$. Without a
substantial change in $1+w_h$ between $z=2$ and $z=0.13$ it is very difficult to obtain a
large change in $\partial\alpha_{em}/\partial\ln(1+z)$ (cf. eq. (\ref{20})) as required for
small $|R|$.
Only for the large value $E=12$ the functional dependence of $\Delta\alpha_{em}(z)$
deviates substantially from a linear behavior in $z$ in the range $0.5<z<2$,
with a strong ``jump'' of $\Delta\alpha_{em}$ by a factor of $5$ between $z=1$ and $z=2$.

Let us next turn to the high values $z\approx 10^{10}$ characteristic for nucleosynthesis.
In our scenario the dominant effect is a change in $m_n/\bar{M}$ which modifies the clock
\cite{CW1}, i.e.
the rate of decrease of temperature in time units $\sim m^{-1}_n$.
A decrease of $m_n$ relative to $\bar{M}$ is equivalent to an increase of $\bar{M}$
at fixed $m_n$ (and thereby to a decrease of Newton's constant). In turn, this decreases
the Hubble parameter for a fixed temperature $T$ (measured in units of $m_n$).
Assuming that the
primordial helium abundance $Y_p$ is in agreement with the value for time independent
couplings within an uncertainty $|\Delta Y_p/Y_p|<8\cdot 10^{-3}$ and using
$\Delta\ln Y_p=\frac{1}{3}\Delta\ln(m_n/\bar{M})$, yields the bounds
$|\Delta\ln(m_n/\bar{M})|<0.025$ or
\begin{equation}\label{A88}
|\frac{\Delta\alpha_{em}(z=10^{10})}{\alpha_{em}}|<0.64\cdot 10^{-3}.
\end{equation}
The effect of the variation of $m_n/\bar{M}$ is much bigger than the influence of the variation
of the proton-neutron mass difference due to electromagnetic effects.
This explains why the bound (\ref{A88}) is more restrictive
than several previous bounds \cite{BBN}, \cite{Av}. We infer from table 2 that our model meets
the bound (\ref{A88}) as well as weaker bounds from the CMB \cite{Av}.
It is striking that the values of
$\Delta\alpha_{em}/\alpha_{em}$ at nucleosynthesis are very far from an extrapolation
with $\partial\ln\alpha_{em}/\partial t=const.$ or even from $\partial\ln\alpha_{em}/
\partial\ln a=const.$.

In our approximation where the dominant field dependence of the various couplings
arises from $\beta_{\alpha}$ we can also estimate the size of the differential acceleration
$\eta$ between two test bodies with equal mass but different composition. One finds
\cite{CWTV}
\begin{equation}\label{22}
\eta=-5\cdot10^{-2}\left(\frac{\beta_{\alpha}}{\alpha_X}\right)^2
\delta^{-1}\Delta R_Z
=-1.75\cdot 10^{-2}\left(\frac{\partial\ln\alpha_{em}}{\partial z}\right)^2_{|z=0}
\frac{\Delta R_Z}{\Omega^{(0)}_h(1+w^{(0)}_h)}
\end{equation}
where for typical experimental conditions $\Delta R_Z=\Delta Z/
(Z+N)\approx 0.1$. Here $\beta_{\alpha}/\alpha_X$ and $\delta$ have to be evaluated at
$z=0$. Results for $\eta$ are also displayed in table 2. We conclude that our CQ-model
is compatible with the present experimental bounds \cite{EP}
$|\eta|<3\cdot 10^{-13}$. We note that for a given value of
$\partial\ln\alpha_{em}/\partial z$ at $z=0$ (e.g. as inferred from Oklo) the differential
acceleration is strongly enhanced for $1+w^{(0)}_h\ll 1$. In addition, the effects of
variations of mass ratios like $M_W/m_n$ or $m_p/m_e$ (that we have neglected here) can
substantially enhance $\eta$ \cite{CWTV}. Already at the present stage the
tests of the equivalence principle can help to discriminate between different models!
Further improvements of the accuracy of tests
of the equivalence principle should either confirm or reject the interpretation of the
QSO-results within our CQ-model.

In summary, we have found models of crossover quintessence that can explain the QSO value
for the time dependence of the fine structure constant and are nevertheless compatible
with all observational bounds on the time variation of couplings and tests of the
equivalence principle. This explanation does not involve any particular cancellation of
effects from the variation of different couplings.
The same models do also very well with all
present cosmological observations!

What will be the tests of such models? First of all, the QSO result should be confirmed by
new independent data. Hopefully this will allow for more precise restrictions on the shape of
the function $\Delta\ln\alpha_{em}(z)$ which can be directly compared with the model predictions.
Second, an improvement of the accuracy of the tests of the equivalence principle
by an order of magnitude should lead to a direct detection of the new interaction mediated
by the cosmon field. In view of its important role in present cosmology we may call this
fifth force (besides gravity, electromagnetism, weak and strong interactions) ``cosmo-dynamics''.
Third, cosmological tests should distinguish crossover quintessence from a cosmological
constant.

The main signature in this respect is perhaps not the equation of state of dark
energy in the most recent epoch (say for $0<z<0.5)$. The value of $w^{(0)}_h$ in table 1
shows that this quantity may be quite near to the value for a cosmological constant (i.e.
$w^{(0)}_h=-1$). Tests of $w_h(z)$ at $z>1$ by supernovae or gravitational
lensing may be more
sensitive. Indeed, if the QSO-results on the time variation of the fine structure constant are
to be explained by quintessence this requires a substantial evolution of the cosmon field in
the epoch near $z=2$ (as compared to the rate of evolution today). We therefore expect that
$1+w_h(z)$ increases with increasing $z$. In turn, this implies that
$\Omega_h(z)$ is substantially larger
during structure formation and last scattering as compared to the value expected for a
cosmological constant. Observations of structure
formation and the CMB may be able to measure the value of $\sigma_8/
\sigma^{(\Lambda)}_8$ and $\Omega^{(ls)}_h$ (table 1) with sufficient precision for
a distinction from the cosmological constant. Finally, an improvement of the
accuracy of atomic clocks by three orders of magnitude could replace the bound
(\ref{2}) by an equivalent bound on
$\partial\ln\alpha_{em}/\partial z$ at $z=0$
within an experimentally well controlled setting.

\end{document}